\documentclass[aps, pra, twocolumn, notitlepage, nofootinbib, showkeys]{revtex4-1}

\usepackage{graphicx,amsmath,pstricks,subfig,float}
\begin{document}
	
\title{Expanding cosmological civilizations on the back of an envelope}
	
\author{S. Jay Olson}
\email{stephanolson@boisestate.edu}
\affiliation{Department of Physics, Boise State University, Boise, Idaho 83725, USA}

\date{\today}
	

\begin{abstract}
We present a simplified description of expansionistic life in the standard relativistic cosmology.  The resulting model is exactly integrable, yielding a simple set of predictive formulas.  This allows one to quickly propose new scenarios for the life appearance rate and the dominant expansion speed and evaluate the observable consequences. These include the expected number and angular size of visible expanding domains, the total eclipsed fraction of the sky, and the life-saturated fraction of the universe.  We also propose a simple anthropic bound on observable quantities, as a function of the dominant expansion velocity alone. The goal is to create a simple and intuition-building tool for use in the context of cosmology, extragalactic SETI, and futures studies.  We discuss the general predictions of this framework, including conditions giving rise to an ``extragalactic Fermi paradox,'' in which zero civilizations are visible beyond the Milky Way. This can occur even if a substantial fraction of the universe is already saturated with ambitious life.
\end{abstract}

\maketitle

\section{Introduction}

\emph{Ambitious} technological life is that which seeks to maximize access to resources.  This appears to be one generic endpoint of technological development, emerging either from selection pressures in a competitive environment of many agents~\cite{hanson1998}, or from a single dominant agent, whose goals can more fully be realized with greater access to resources~\cite{bostrom2014}.  If a few technologies are practical (e.g.\ high-speed~\cite{merali2016}, long-range~\cite{fogg1988,armstrong2013} spacecraft with self-reproducing capability~\cite{freitas1980}), an ambitious civilization would manifest in an expanding domain of influence that could reach a cosmological scale after sufficient time~\cite{olson2014}.  Due to the exponential reproduction implied by self-reproducing technology, such an effort is not necessarily expensive (the total cost need not be more than one self-replicating spacecraft), and the required number of generations of replication is not extreme.  We have thus proposed searching for cosmic-scale domains consisting of many life-altered galaxies, as an approach to extragalactic SETI~\cite{olson2016,olson2017a}.

One barrier to this kind of SETI project is the theory required to characterize such targets in a wide range of scenarios.  It combines relativistic optical effects (in a background FRW spacetime) and anthropic information~\cite{olson2017a,olson2016}.  This is a virtual guarantee that some predictions will be at odds with intuition.  To complicate matters, previous theoretical work has emphasized general expressions with many parameters, evaluated numerically, which creates another barrier to intuition and use.

Our purpose here is to correct this situation by introducing approximations to reduce the number of parameters and to make the theory analytically integrable, reducing the core predictions to set of simple functions.  The end result retains the essential (sometimes counter-intuitive) features of the full version, but without the requirement of numerical integration.

Section II reviews the background and introduces the basic approximations to be made.  Essentially, we reduce the number of parameters in the model, and approximate an integral that appears in all predictive expressions.  Section III gives the predictive expressions as functions of this integral and two parameters (the life appearance rate $\alpha$ and dominant expansion speed $v$).  These quantities include e.g.\ the life-saturated fraction of the universe, the expected number and angular size of visible domains, the probability that at least one domain is visible, and the total expected fraction of the sky eclipsed by expanding domains.  New scenarios can be invented and the observable consequences calculated immediately.  We also introduce correlations between these quantities that are obtained by eliminating dependence on $\alpha$, e.g.\ we can express the probability of seeing at least one domain as a function of $v$ and the currently life-saturated fraction of the universe $h(t_0)$.

Section IV introduces a simple anthropic bound on the appearance rate constant $\alpha$, for which humanity is in the final $\approx 7\%$ of civilizations to appear in humanity-like conditions.  The main advantage of this bound is that it simplifies predictions still further, reducing the predictive quantities to elementary functions of $v$ alone. 

Section V discusses implications of these results.  For example, we examine sensitivity to the highly-uncertain value of $\alpha$, and conclude that unless it is simultaneously true that $\alpha$ is large (very close to the anthropic bound) \emph{and} $v$ is not in the high speed regime, an extragalactic Fermi paradox appears as the default expectation.  We also discuss implications for searches of isolated Kardashev type iii (K3) galaxies~\cite{kardashev1964}.  Unless K3s appear at a rate that is orders of magnitude greater than ambitious civilizations, K3s are not expected to be found closer than fully cosmological distances and thus will be difficult to detect.  Section VI contains our concluding remarks.  

Throughout this paper, we use the convention that $c=1$ to simplify expressions and so that velocity (in the co-moving reference frame) is always expressed as a number less than one.  This implies Gly$=$Gyr for our distance/time units of choice, though we make use of both to distinguish space from time.

\section{Background and approximations}

A relativistic cosmology with ambitious life is based on a description of random \emph{appearance events} in spacetime, where each event gives rise to high-speed spherical expansion, very analogous to a cosmological phase transition~\cite{olson2014}.  The appearance rate per unit co-moving volume, $f(t)$, is expressed as $f(t) = \alpha F(t)$, where $F(t)$ encodes the relative time dependence, normalized to a maximum value of unity, so that $\alpha$ is interpreted as the \emph{peak} appearance rate of such life in the universe.  This is useful because reasonable estimates for $F(t)$ can be constructed, based on assumptions about the cosmic time-dependence of habitable conditions (e.g. planet formation rates and stellar lifetimes) and the apparent timescale of biological evolution, while the extreme prior uncertainty of the appearance rate can be confined to the parameter $\alpha$.

After an appearance event at time $t_i$ has taken place, the phase of rapid spherical expansion is described in terms of the volume occupied at time $t$, given by $v^3 V(t_i,t)$, where $V(t_i,t) = \frac{4 \pi}{3} \left( \int_{t_i}^{t} \frac{1}{a(t')} dt' \right)^3 $ is the co-moving 3-volume of a slice of the future light cone at future time $t$.

The framework studied up until now has made use of five parameters per type of expanding civilization one wishes to include in the model, $ \{ \alpha, v, T, A, \Gamma \} $, representing the appearance rate, the expansion speed, the time lag between arrival at a galaxy and observable changes to the galaxy, the fraction of baryonic matter available to be used as fuel, and the rate of mass conversion to waste radiation.  Most of these parameters are of purely theoretical interest.  As a tool for making SETI predictions, $T$, $A$, and $\Gamma$ are unlikely to change the basic conclusions of any realistic model.  For example, a very large $T$ induces a cosmic time-dependence in the expansion velocity of the visible boundary of a domain, but even if $T$ is $100$ million years (an extremely long time for galaxy colonization compared to long-established theoretical timescales~\cite{jones1976}), the effect will amount to less than a one percent shift in expansion velocity over nearly the entire duration in which expansionistic life could have been active.  The mass availability and consumption rate are interesting in the sense that they determine the backreaction of life on the evolution of the universe~\cite{olson2014}, but such a backreaction is easily small enough to be ignored for the purposes of SETI.

It is also reasonable to expect that nearly all ambitious civilizations will expand at close to the same velocity.  This can be motivated by observing that the cost of spacecraft speed must rise sharply within a narrow range of $v$.  At $v=.2$, interstellar probes are already regarded as practical and are being actively researched in the present-day Earth economy~\cite{merali2016}, but the cost must become infinite for any civilization at some (unknown) speed strictly less than 1.  Thus, all technologically mature civilizations with advanced energy resources will quickly discover the same (approximate) practical speed limit, whatever it may be.  If such a civilization is \emph{ambitious}, seeking to acquire maximum resources, their net expansion speed will be close to this limit.  This means that a model including only a single expansion velocity parameter, $v$, common to all ambitious civilizations, is a plausible assumption for extragalactic SETI.

We are now down to two total parameters $\{ \alpha, v \}$ and a life appearance rate model $F(t)$ to be specified.  In section IV, we will introduce an anthropic bound that takes the simplification even further, requiring only a single estimate of $v$ for the purposes of setting bounds on predictions.  Section III will give more general expressions that require $\{ \alpha, v \}$ and $F(t)$ to be specified.

The form of $F(t)$ is periodically discussed in the literature~\cite{lineweaver2001,loeb2016,olson2017a}, and at late cosmic times, $F(t)$ will depend heavily on the assumed habitability of different star types~\cite{loeb2016}.  If one assumes that practically all intelligent life will appear around sun-like stars, then $F(t)$ will have mostly died out when the universe is twice its current age.  If one assumes that intelligent life is most likely to appear around m-dwarfs at any time throughout their long lives, then $F(t)$ will be approximately flat for another ten trillion years.  Here, we are interested in exact integration, so we use piecewise linear approximations to $F$.  Two illustrative examples are:

\begin{displaymath}
F_1(t) = \left\{
\begin{array}{lcc}
 \frac{(t- 7)}{4}  & : &   7  < t < 11 \\
1-\frac{t-11}{15} \,   & : &  11 < t < 26
\end{array}
\right.
\end{displaymath} 
  
\begin{displaymath}
F_2(t) = \left\{
\begin{array}{lcc}
\frac{(t- 7)}{4}  & : &   7  < t < 11 \\
1 \,   & : &  11 < t 
\end{array}
\right.
\end{displaymath} 

Figure 1 compares $F_1$ to a previously used~\cite{olson2014,olson2017a,olson2017b}, numerically integrated model that we denote here as $F_{num}$.  $F_{2}$ is designed to be a rough approximation to a dominant m-dwarf model.

\begin{figure}
	\centering
	\includegraphics[width=0.9\linewidth]{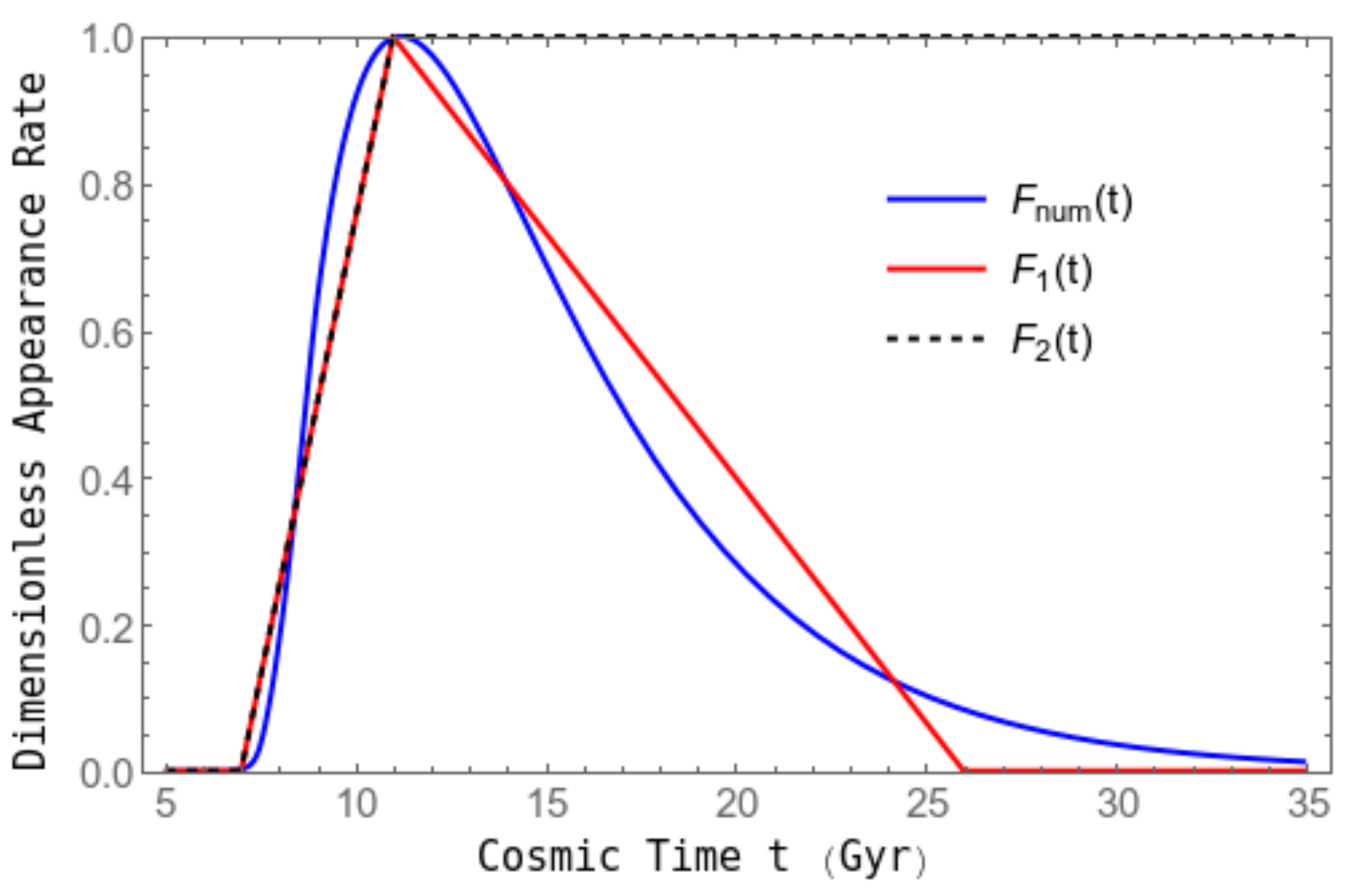}
	\caption{Three models of the time dependence $F(t)$ of the appearance rate of expansionistic life at the cosmic scale. $F_{num}(t)$ is a numerically integrated model that tracks the formation rate of earth-like planets around sun-like stars, with a lag of several Gyr to allow for biological evolution.  $F_1 (t)$ is a piecewise linear approximation to $F_{num} (t)$, and $F_2 (t)$ is rough alternative model for which planets around m-dwarfs are the dominant source of advanced life.}
\end{figure}

To obtain predictions, the chosen $F(t)$ must be integrated within our past light cone, so that the following integral (denoted here as $s(t_0)$) will appear in every single predictive expression:  
  
\begin{eqnarray}
s(t_0) = \int_{0}^{t_0} F(t) V(t,t_0) \, dt.
\end{eqnarray}

To obtain $s(t_0)$ by analytic integration, we can expand $V(t,t_0)$ as a power series around $t_0$:
\begin{widetext}
\begin{eqnarray} 
	V(t,t_0) \approx \frac{4 \pi}{3} \left(-(t-t_0)^3 + \frac{3}{2} H_0 (t-t_0)^4 +\frac{1}{4} \left(-2 H_0^2 q_0 - 7 H_0^2\right) (t-t_0)^5 \right) + O((t-t_0)^6).
\end{eqnarray}
\end{widetext}
Here, $H_0$ is the Hubble constant, and $q_0$ the deceleration parameter\footnote{In all calculations, we assume the following cosmology:  $\Omega_{\Lambda 0}=.692$, $\Omega_{r0}=9 \times 10^{-5}$, $\Omega_{m0}=1-\Omega_{r0} -\Omega_{\Lambda 0}$, $H_0 =.069 \, Gyr^{-1}$, giving $t_0 = 13.8$ Gyr and $q_0 = \frac{\Omega_{m 0}}{2} + \Omega_{r 0} - \Omega_{\Lambda 0} = -.538$.}.  In the above, we have implemented $a(t_0)=1$ to keep the expression as clean as possible.

The full numerical version gives $s(t_0) = 1268 \, Gly^3 \, Gyr$, while the $F_1$ approximation (with 5th order approx. to $V(t,t_0)$) gives $1223 \, Gly^3 \, Gyr$ and $F_2$ gives $1226 \, Gly^3 \, Gyr$.  One can see that the distinction between $F_1$ and $F_2$ will be unimportant for observers at $t_0$ -- the very different physical assumptions will result in substantially different predictions only at much later cosmic times. 

Using $F_1(t)$ and the full 5th order approximation to $V(t,t_0)$ give $s(t_0)$ to within 3.5\% of the full numerically integrated result\footnote{Manually evaluating $s(t_0) = \int_{0}^{t_0} F_{1}(t') \, V(t',t_0) \, dt'$ in this way amounts to a repeated application of $\int (t-a)^{n} \, (t-b) \, dt = \frac{(t-a)^{n+1} (a-b (n+2)+n t+t)}{(n+1) (n+2)}$ for each power in the power series approximation of $V(t,t_0)$.}, while using the 4th order approximation to $V(t,t_0)$ yields an approximation that is good to 13\%.  Keeping only the 3rd order term yields a 41\% error.  Errors in this approximation will naturally grow when models for $F(t)$ are chosen that put greater weight at early times, i.e. farther from $t_0$.

\section{Predictive quantities of interest}

With $s(t_0)$ approximated according to the previous section, a number of quantitative predictions can be made, as a function of parameters $\alpha$ and $v$.  The general results of our previous work reduce, in the above approximation, to simple, easily calculable expressions.  

The fraction of space that is \emph{not} saturated with ambitious life has the form $g(t)=e^{-\alpha v^3 s(t)}$, so the unsaturated fraction of the universe at the present time is~\cite{olson2014}:
\begin{eqnarray}
g(t_0)=e^{-\alpha v^3 s(t_0)}.
\end{eqnarray}

Our analysis is based on the core assumption that appearances of expansionistic life are random and independent events in the universe, i.e. a Poisson process, so that the expected (average) number of expanding domains that appear within our past light cone (i.e. are visible to us\footnote{We must mention a potential caveat.  We are assuming that such civilizations are visible if they appear within our past light cone, but one can imagine possibilities in which they would not be observable, e.g. if they wish to avoid detection or are interested in spreading out over the cosmos but not interested in making use of resources in a way that would be visible at the present time.  It is not difficult to extend the analysis to include such cases, but it is beyond our present scope.}), $E(n)$, is~\cite{olson2017a}:
\begin{eqnarray}
E(n)=\alpha (1-v^3) s(t_0),
\end{eqnarray}
and in general, the probability that $n$ are visible is thus:
\begin{eqnarray}
p(n) = \frac{e^{- \alpha \, s(t_0) \left(1-v^3\right)} \left(\alpha \, s(t_0) \left(1-v^3\right)\right)^n}{n!}.
\end{eqnarray}

Thus, the probability of \emph{at least one} domain being visible to us is:

\begin{eqnarray}
p(n \geq 1) = 1 - e^{-\alpha (1-v^3) s(t_0)}.
\end{eqnarray} 

The recurring factor of $1-v^3$ can be understood as multiplying $V(t',t)$ within $s(t_0)$ to give the volume in our past light cone \emph{minus} the volume in our ``past saturation cone'' (emanating into our past with speed $v$) that we know to be empty, else our galaxy would already be saturated.

The formulas for $E(n)$ and $p(n)$ are approximations that do not properly exclude the possibility of expanding domains appearing within already expanding domains, but they should remain useful, so long as $n$ is not extremely large.  The expressions for $g(t)$ and $p(n \geq 1)$ do not suffer from this defect.

The expected fraction of the sky for which the line of sight does \emph{not} intersect a visible expanding domain at any distance, is given by $k$~\cite{olson2016}:
\begin{eqnarray}
k = e^{-\frac{\alpha v^2 (1-v)^2}{4} s(t_0)}.
\end{eqnarray}

To estimate the average angular size (expressed as a fraction of the celestial sphere) of a visible expanding domain, we can divide the fraction of the sky expected to contain occupied galaxies, $(1-k)$, by the expected observable number, $E(n)$:

\begin{eqnarray}
E(\Omega)=\frac{(1-k)}{E(n)}.
\end{eqnarray}

Again, $E(\Omega)$ is an approximation that does not account for the fact that domains might eclipse one another, but it will remain useful so long as the expected fraction of the sky eclipsed by \emph{zero or one} expanding civilization is dominant, i.e. that $(1+\frac{\alpha v^2 (1-v)^2}{4} s(t_0))e^{-\frac{\alpha v^2 (1-v)^2}{4} s(t_0)}$ is not too far from 1, following the Poisson property. 

One last result that hints in the direction of anthropic reasoning comes from eliminating dependence on $\alpha$ and $s(t_0)$ and expressing $p(n \geq 1)$ (probability that at least one ambitious civ.\ is visible) as a function of $v$ and $h(t_0) = 1 - g(t_0)$ (i.e.\ the fraction of the Universe that is saturated by life at the present time).  The result is:

\begin{eqnarray}
p(n \geq 1) = 1- (1-h)^{v^{-3} -1}.
\end{eqnarray}

This function is depicted in figure 2, for several values of $v$.  One can immediately see that conditions must be extreme for any hope of detection, if the dominant expansion velocity $v$ is high, i.e. the universe must already be mostly saturated if we expect to see anything.  This formula is particularly general, since it is independent of the appearance rate model and cosmological parameters.  The same formula must be true for \emph{any} civilization analogous to humanity appearing at any time in the history or future of the universe.  One can, of course, express the other quantities in terms of one another as well, leading to similar general expressions that eliminate dependence on the parameters.

\begin{figure}
	\centering
	\includegraphics[width=0.9\linewidth]{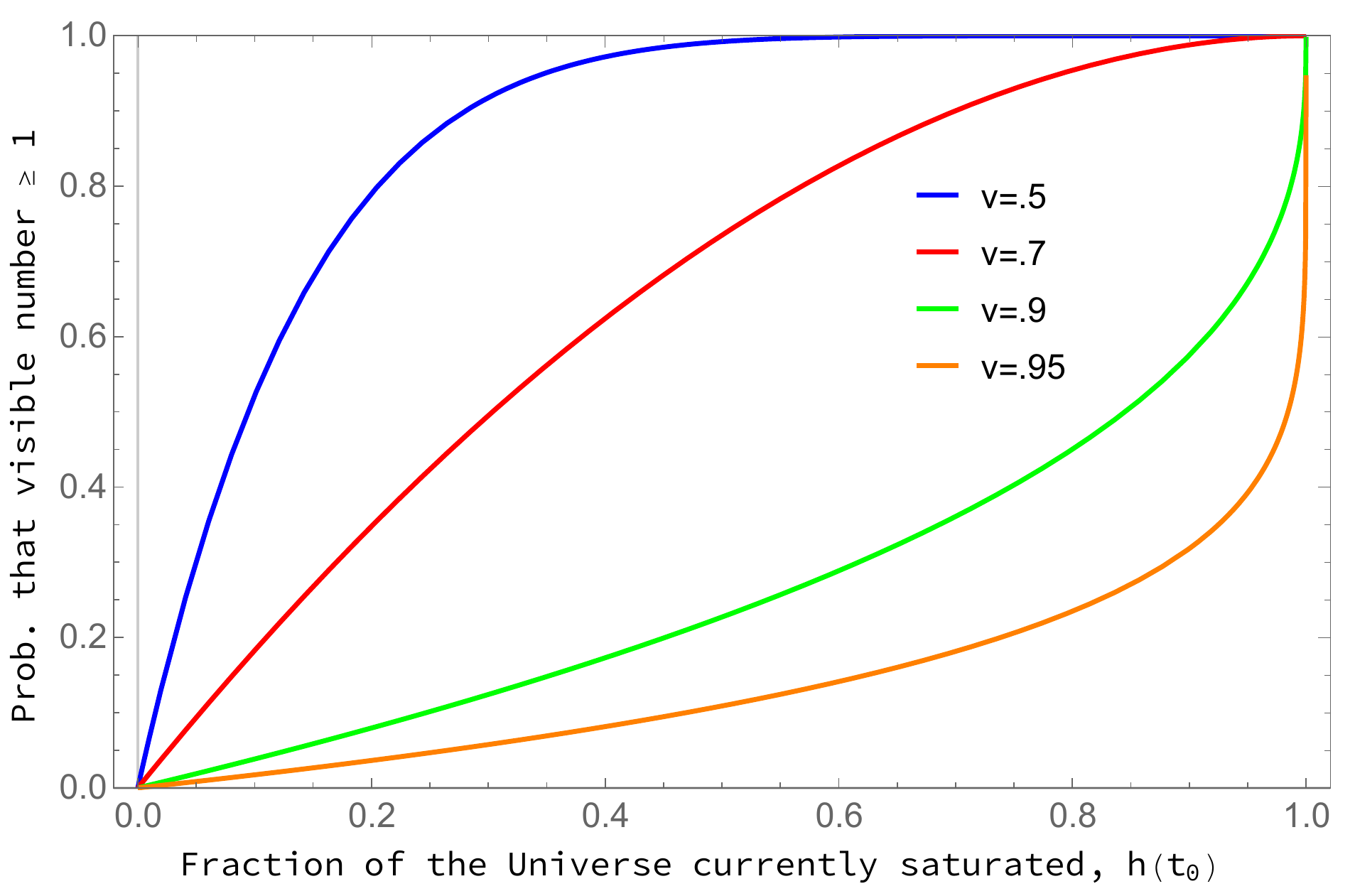}
	\caption{Probability that at least one domain is visible, as a function of the currently-saturated fraction of the Universe, $h(t_0)$, for several values of the dominant expansion velocity (equation 9).  In high-$v$ scenarios, one cannot expect to see expanding domains unless the universe is already heavily saturated with ambitious life.  This prediction is quite general as it is independent of models for the appearance rate.}
\end{figure}

\section{A simple anthropic bound}
 
The fundamental difficulty in making predictions is that, while $v$ and $F(t)$ can reasonably be estimated from independent information, the value of $\alpha$ is radically uncertain over many orders of magnitude~\cite{tegmark2014,lacki2016}.  To constrain the possibilities, we would like a condition that tells us when estimates for $\alpha$ are unreasonably large.  

Up to now, this has been done by assuming that humanity is fairly typical in the set of comparable, human-stage civilizations who have arrived to find themselves in a galaxy that appears untouched by ambitious life~\cite{olson2017a}.  In particular, our cosmic time of arrival should not be an extreme outlier in this set. 

If we assume that human-stage civilizations appear with the same cosmic time-dependence $F(t)$ as ambitious civilations (but with a strictly larger, though unknown proportionality constant), then the time of arrival of a random human-stage civilization is drawn from the following probability density function:

\begin{eqnarray}
p_{TOA}(t)=\frac{g(t)F(t)}{\int_{0}^{\infty} g(t') F(t') \, dt'}.
\end{eqnarray}

Note that this pdf does not depend on the proportionality constant for human-stage life (it would appear in both the numerator and denominator, factoring out), but it does depend on $\alpha$ (through $g(t)$), because the ambitious civilizations displace the untouched galaxies with time.  If $\alpha$ is too large, causing $g(t)$ to decreases too rapidly in cosmic time, then we humans at cosmic time $t_0$ have arrived at an implausibly late time in $p_{TOA}(t)$.

Using a numerical approach, the standard way to set an anthropic bound on $\alpha$ is to search for the value of $\alpha$ that begins to make appearing at $t_0$ implausible.  We could find, for example, the value of $\alpha$ such that $t_0$ is two standard deviations from the mean time of arrival, or if humanity has appeared in the final few percent of human-stage civilizations~\cite{olson2017a}. 

For back-of-the-envelope purposes, we take an alternate approach; we seek an easy-to-solve condition on the parameters, stating that opportunities for life to appear in human-like conditions are rapidly coming to an end.  The condition we advocate here is to choose the value of $\alpha$ to maximize the magnitude of $\frac{d g(t)}{dt}$ at $t_0$.  We denote this as the ``maximum g-slope" (mgs) condition.  This condition selects the following value of $\alpha$:

\begin{eqnarray}
\alpha_{mgs} = \frac{1}{v^3 \, s(t_0)}.
\end{eqnarray}

This also implies that $g(t_0)=e^{-1}$, i.e. that $\approx 63 \%$ of galaxies in the universe are currently saturated with ambitious life.  This may sound extreme, a condition easily falsified by simple observation, but the previous section shows that observability depends heavily on $v$.  If $v$ is high enough, not a single saturated galaxy is expected to be observable under the mgs condition.  It is immediate to evaluate the relevant quantities from the previous section for the mgs condition.

Expected visible number:
\begin{eqnarray}
E_{mgs}(n)= v^{-3} - 1.
\end{eqnarray}

Probability that $n$ are visible:
\begin{eqnarray}
p_{mgs}(n) = \frac{e^{-\left(v^{-3}-1\right)} \left(v^{-3}-1\right)^n}{n!}.
\end{eqnarray}

Probability that at least one is visible:
\begin{eqnarray}
p_{mgs}(n \geq 1)=1-e^{-(v^{-3}-1)}.
\end{eqnarray}

Expected fraction of sky not eclipsed: 
\begin{eqnarray}
k_{mgs} = e^{-\frac{(1-v)^2}{4 v}}.
\end{eqnarray} 

Average angular size of visible domains (as a fraction of the celestial sphere): 
\begin{eqnarray}
E_{mgs}(\Omega)=\frac{v^3 \, \left( 1 - e^{-\frac{(1-v)^2}{4 v}} \right)}{1 - v^3}.
\end{eqnarray}
 
We now have an anthropic bound on $\alpha$, giving rise to a set of observable quantities that are reduced to elementary functions of a single parameter, $v$.  The value of $v$ is in principle a technological question, and is uncertain to less than a single order of magnitude.  The predictions of the mgs condition are illustrated in Figure 3:

\begin{figure}
	\centering
	\subfloat[]{
		\includegraphics[width=0.5\linewidth]{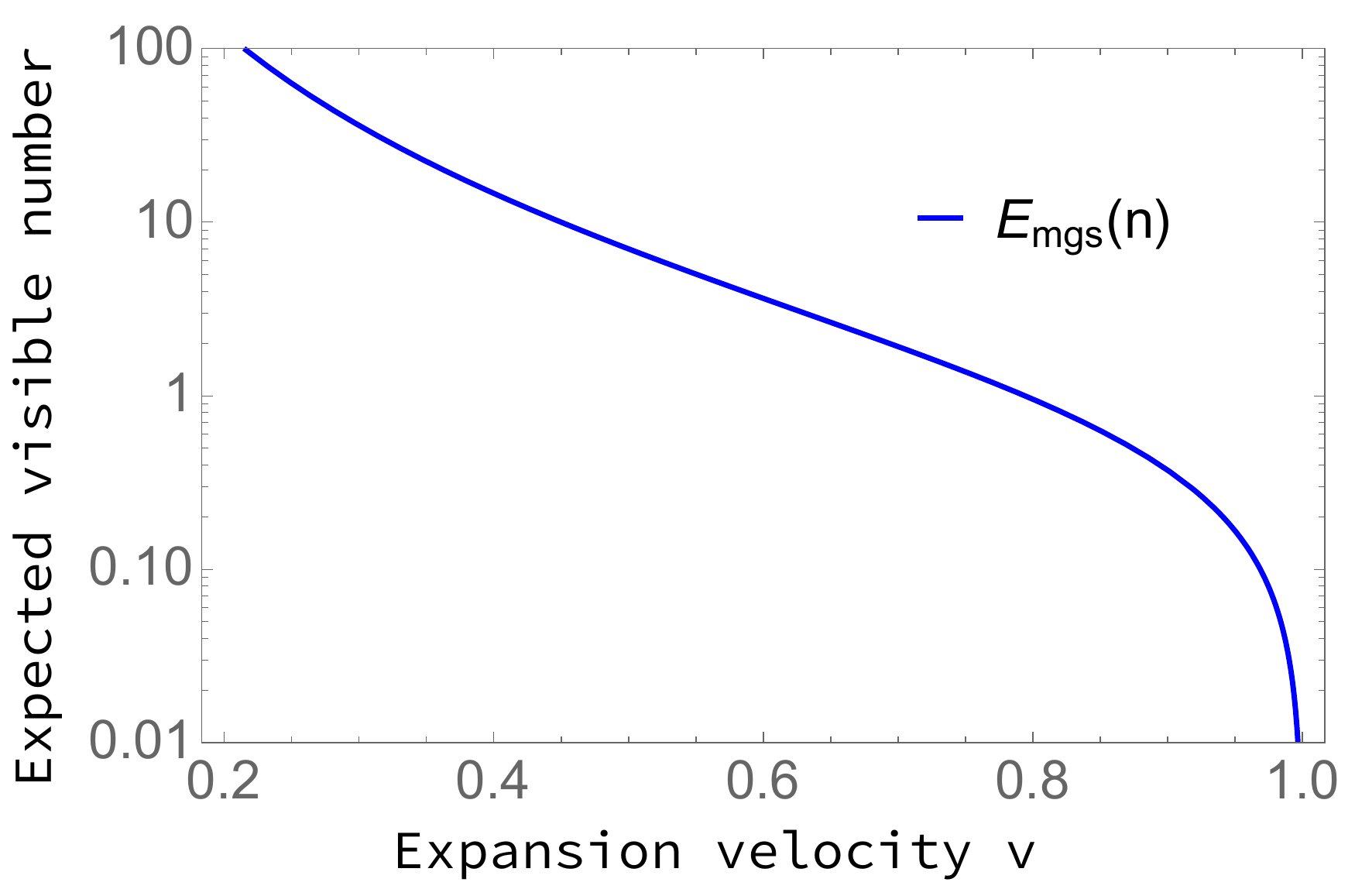}
	}
	\subfloat[]{
		\includegraphics[width=0.5\linewidth]{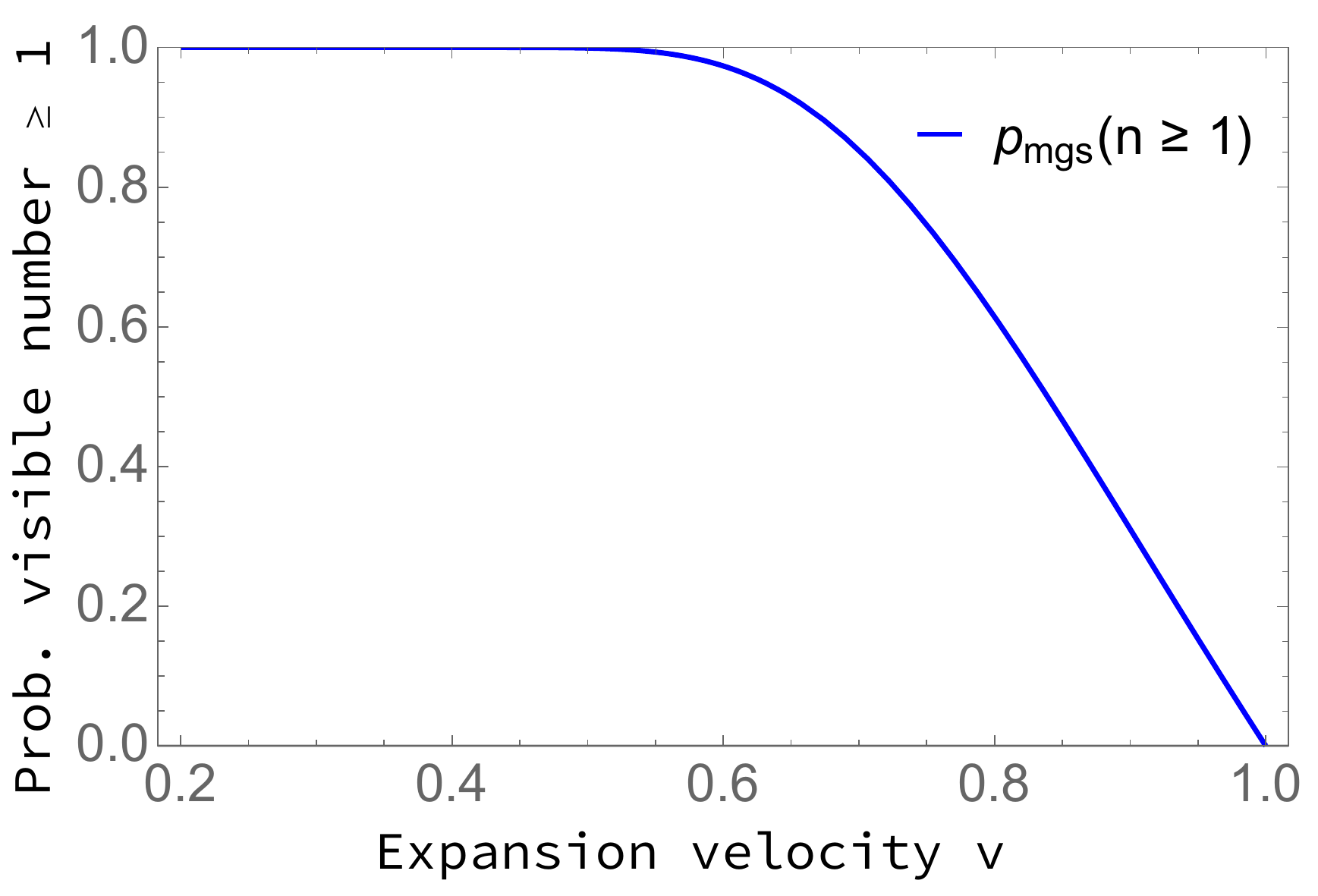}
	}
	\hspace{0mm}
	\subfloat[]{
		\includegraphics[width=0.5\linewidth]{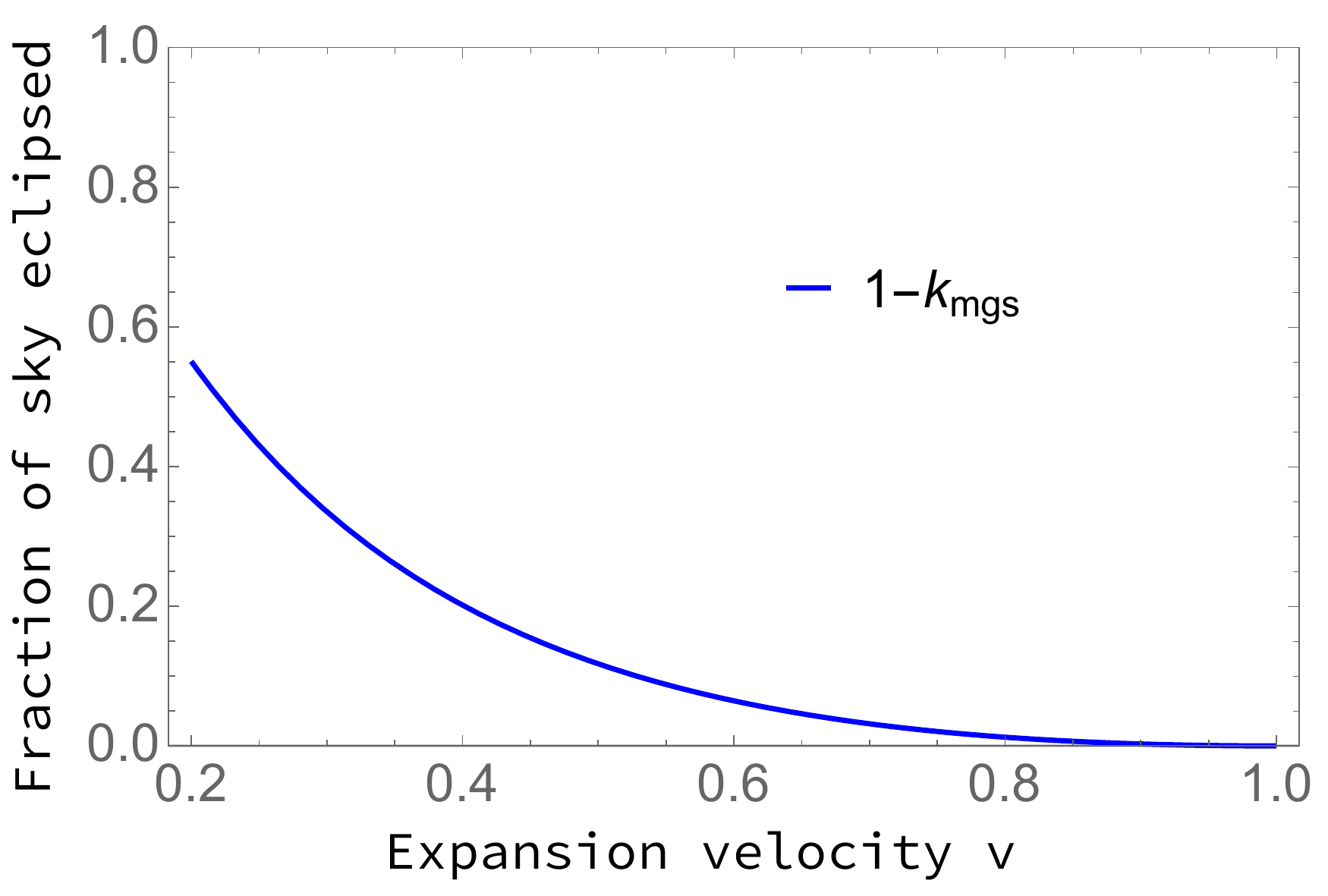}
	}
	\subfloat[]{
		\includegraphics[width=0.5\linewidth]{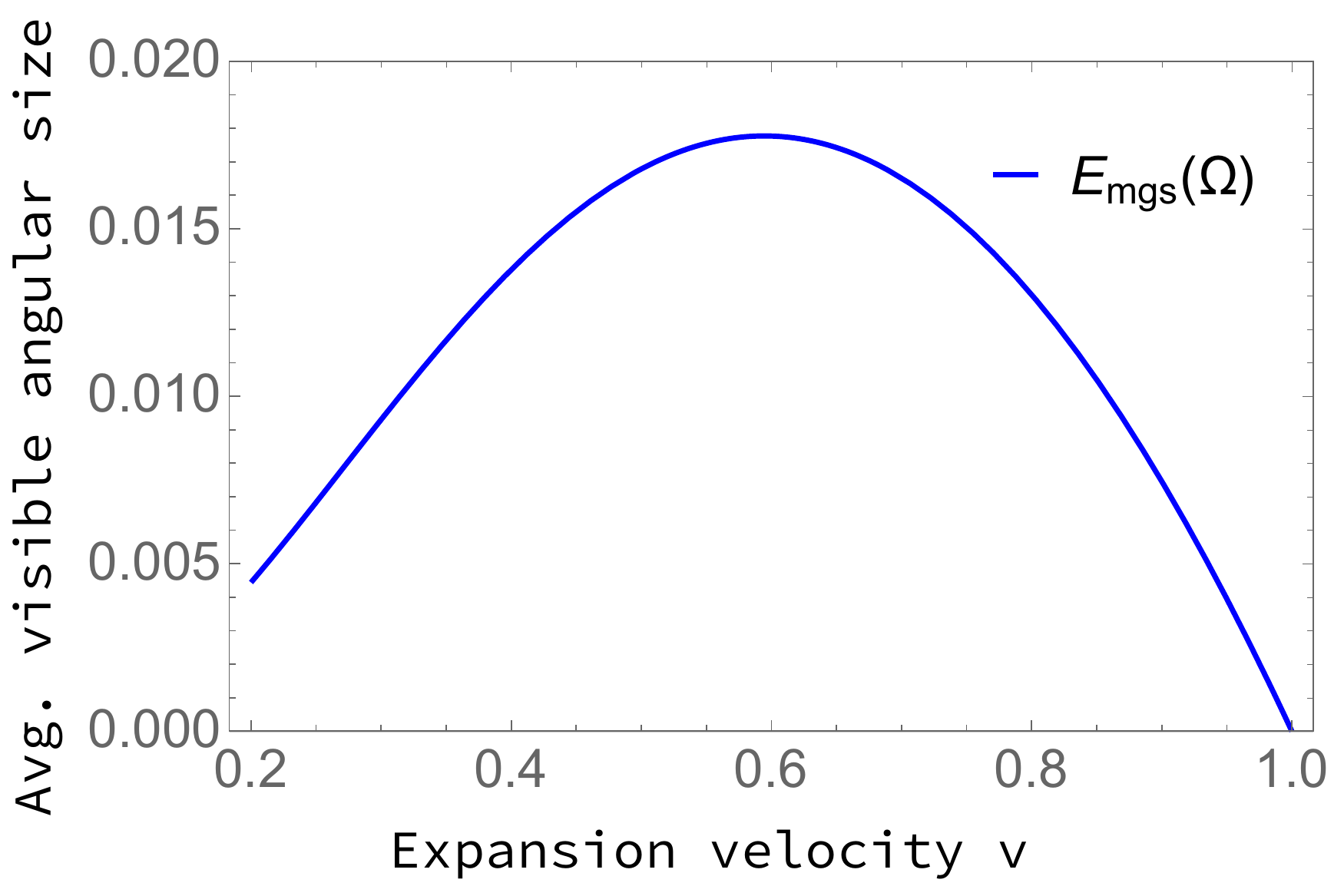}
	}
	\caption{Predictions of the maximum g-slope (mgs) anthropic condition on the appearance rate (in which we are in the final $\approx$ $7 \%$-$10 \%$ of human-stage civilizations to appear in humanity-like conditions), as a function of the dominant expansion velocity of ambitious civilizations in the universe.  }
\end{figure}

A notable feature of this bound is the way it separates the role of the parameters $\{ \alpha, v \}$ from the role of the appearance rate model $F(t)$.  The bound gives expressions that depend only on $v$, eliminating dependence on $F(t)$.  On the other hand, the \emph{strength} of the bound is determined entirely by the appearance rate $F(t)$ alone, i.e. it is parameter-independent.  This can be seen by noting that all parameter-dependence disappears from $P_{TOA}(t)$, when $\alpha_{mgs}$ is substituted into $g(t)$.  This makes the mgs bound extremely convenient for some purposes, but unsuitable for others.

The mgs condition puts us within the final $\approx 7\% -10 \%$ of human-stage civilizations appearing in human-like conditions, for many plausible choices of $F(t)$.  In particular, $F_{num}(t)$ puts us in the final $7.2 \%$, $F_1(t)$ the final $8.4 \%$, and $F_2(t)$ the final $ 10.4 \% $ of civilizations in humanity-like conditions.  It is difficult to find a \emph{realistic} $F(t)$ for which the mgs condition fails to be a reasonable (if not particularly strong) anthropic bound on $\alpha$, for which our prior uncertainty covers many orders of magnitude.

\section{First consequences and extensions}

We are now in a position to examine some general conclusions for extragalactic SETI.  Some have been arrived at before~\cite{olson2016,olson2017a}, but their origin should now be more immediately clear, and they can be extended.  We consider in this section the sensitivity of observable quantities to the value of $\alpha$ when incorporating prior uncertainty and the anthropic bound, and we extend our results to non-expanding type iii galaxies, and the expectation of an extragalactic Fermi paradox.

\subsection{Sensitivity to the value of $\alpha$}

The mgs condition gives predictions based on the largest plausible value of $\alpha$, while section III tells us how observability scales with $\alpha$.  For example, the mgs bound on $E(n)$ (the average number of visible domains) is less than ten over the window of $.5 < v < 1$ (and less than 1 for $v>.8$), and $E(n)$ scales linearly with $\alpha$.  Thus, there exists a very narrow window in the value of $\alpha$ that would lead to likely observable expanding civilizations.  If the actual value of $\alpha$ is not within a single order of magnitude of the mgs anthropic bound, present-day humans will not be observing ambitious civilizations in the high-speed regime.

The low-speed regime of $.2 < v < .5$ extends the window of observability in $\alpha$ to one or two orders of magnitude from the mgs bound.  While this is still quite a narrow region for such an uncertain parameter, we also run into the seemingly unlikely prospect of Kardashev type iii+ civilizations limited to spacecraft speeds very similar to those currently under active development on Earth.

One can make this more quantitative by assigning a prior probability distribution over $\alpha$, reflecting the state of our ignorance.  One approach is to assign a prior pdf that is constant in $\log(\alpha)$, from some lower cutoff $\alpha_{min}$, up to the mgs value of $\alpha$ at the high end~\cite{tegmark2014}.  This will depend quite heavily on the cutoff -- if one chooses e.g. $\alpha_{min}=10^{-1000}$, then the prior probability to be within one order of magnitude of the mgs value (for e.g. $v=.5$) will be roughly a tenth of a percent.  Prospects for observation get correspondingly better or worse, depending on the chosen exponent in the cutoff.

Another approach is to choose a prior that is constant in $\log (\log (\alpha) )$.  Use of this prior reflects a situation in which we are also highly uncertain of the number of important-but-rare environmental factors that must be present for the successful evolution of intelligence~\cite{lacki2016}.  This prior is less sensitive to the lower cutoff, and tends to assign more weight to the observable range when the cutoff value for $\alpha$ is tiny.  With a lower cutoff $\alpha_{min}$ ranging from $10^{-100000}$ to $10^{-100}$, the prior probability of being within an order of magnitude from the mgs bound (for $v=.5$) ranges from $\approx 3 \%$ to $\approx 10 \%$, yielding surprising consistency.  This probability is also similar to the strength of the mgs bound itself, i.e. we must remember that there is a prior probability of a few percent that we have arrived relatively later in $p_{TOA}(t)$ than assumed by the mgs condition.

\subsection{Non-expanding type iii civilizations}

All of the foregoing can be used directly in evaluating searches for expanding cosmological civilizations.  Naturally, there will always be additional constraints in a realistic search, e.g. that most searches cannot cover the full sky, and that the most likely signature of saturated galaxies is debatable.

One important point is that most attempts at intergalactic SETI so far have focused on searching for isolated Kardashev type iii galaxies, i.e. a civilization that underwent exponential expansion to $\approx 10^{11}$ stars, and then permanently stopped at the boundary of its home galaxy.  Let us describe the appearance rate of such civilizations as $\beta F(t)$, where $F(t)$ is the same time dependence as that of expanding civilizations, but $\beta$ is distinct from $\alpha$.  The expected number of those visible to us, in principle, will be $E(n)=\beta s(t_0)$, though an isolated type iii galaxy would presumably be far more difficult to identify at a distance of multiple Gly, in comparison with an extended region of modified galaxies.

An immediate question is whether we can put any constraints on the magnitude of $\beta$, since such civilizations do not displace enormous amounts of space, and the anthropic argument we used to bound $\alpha$ will not be applicable.  The approach so far~\cite{olson2017a} has been to note that expanding to the boundary of a home galaxy and then stopping is unusual behavior.  Recent work has pointed out that technological requirements for intergalactic spacecraft are nearly the same as for interstellar spacecraft~\cite{armstrong2013}, so technological limitations are unlikely to dictate this behavior.  Furthermore, additional resources can be put to use to advance nearly any final goal~\cite{bostrom2014} -- if galaxy-scale expansion is deemed beneficial, then it is difficult to imagine how intergalactic expansion will fail be deemed beneficial, in a generic case.

These considerations lead to the \emph{expansion hypothesis}~\cite{olson2017a}, that the value of $\beta$ should not be orders of magnitude greater than the value of $\alpha$.  This is a conservative statement -- in fact the underlying argument suggests that $\beta << \alpha$.  Nevertheless, if we take as a bound $\beta_{max} = \alpha_{mgs} = \frac{1}{v^3 \, s(t_0)}$ (our anthropic bound on $\alpha$, for a given $v$), then we estimate there to be $E(n) = v^{-3}$ isolated type iii's in our past light cone, at maximum.  Note that $v$ refers here to the speed of the expanders (who determine the mgs condition), and not the isolated type iii's, who do not expand at all on a cosmic scale.

Thus, including the possibility of isolated type iii civilizations seems not to change the basic conclusions coming from the analysis of expanding cosmological civilizations. Again the entire high-speed regime of $v>.5$ is limted to less than $10$ isolated type iii's in our past light cone, and the low-speed regime could contain as many as $\approx 100$.  All would almost certainly appear at cosmological distances, since $s(t_0)$ is dominated by the volume of space at large distance.  As before, this scenario is just a conservative upper bound -- the prior over the appearance rate will dictate the probability that $\beta$ is close to this maximum value.

\subsection{An Extragalactic Fermi Paradox}

Essentially, the default expectation is that zero modified galaxies will be visible from our present vantage point -- to search will be to hope for exceptionally good luck.  Although previous works described the possibility of intergalactic spacecraft in terms of exacerbating the Fermi paradox~\cite{fogg1988,armstrong2013}, our analysis gives a different picture.  We are almost guaranteed to encounter an extragalactic Fermi paradox unless $\alpha$ is so large that the universe is already substantially saturated with life, and in the context of a low-$v$ expansion scenario.  An argument that such a scenario is \emph{likely} would again change the picture, but a large prior uncertainty in $\alpha$ and the likelihood that superintelligence will be able to expand at high-$v$ tends to drive our expectations in the direction of a Fermi paradox (i.e. zero detections, no matter how carefully we look).

\section{Concluding remarks}

An important aspect of this work is that it highlights the optimist/pessimist boundary for extragalactic SETI.  One need only specify an opinion of a single number --- the dominant expansion velocity in the universe, $v$ --- to begin to making an estimate on the likelihood of success for such projects (via the mgs condition of section IV), and the likely observable signatures.  It is also quite easy to take the next step and make more detailed arguments regarding the uncertainty of the appearance rate parameter, $\alpha$, and/or the relative appearance rate $F(t)$ (using section III).  None of this requires numerical integration -- at most, a single analytic integral can be performed by hand.

Beyond these results, there are various possible behavior patterns that might be invoked to modify the basic predictions, reminiscent of many attempts to solve the classical Fermi paradox in the 20th century~\cite{webb2002} (e.g.\ they could deliberately be expanding silently).  It is reasonable to consider such possibilities, but our results show that merely taking relativity into account already imposes serious limitations on visibility, at the cosmic scale.  For an earth-based observer embarking on extragalactic SETI, the most immediate question is much less ``where is everybody?''~\cite{Jones1985} and much more ``do I feel lucky?''

\bibliography{ref5}{}

\begin{thebibliography}{10}

\bibitem{armstrong2013}
Stuart Armstrong and Anders Sandberg.
\newblock Eternity in six hours: Intergalactic spreading of intelligent life
  and sharpening the fermi paradox.
\newblock {\em Acta Astronautica}, 89:1--13, 2013.

\bibitem{bostrom2014}
Nick Bostrom.
\newblock {\em Superintelligence: Paths, dangers, strategies}.
\newblock Oxford University Press, 2014.

\bibitem{fogg1988}
Martyn Fogg.
\newblock Feasibility of intergalactic colonisation and its relevance to setl.
\newblock {\em Journal of the British Interplanetary Society}, page 491, 1988.

\bibitem{freitas1980}
Robert~A Freitas~Jr.
\newblock A self-reproducing interstellar probe.
\newblock {\em Journal of the British Interplanetary Society}, 33:251--264,
  1980.

\bibitem{hanson1998}
Robin Hanson.
\newblock Burning the cosmic commons: Evolutionary strategies for interstellar
  colonization.
\newblock {\em preprint available at http://hanson.gmu. edu/filluniv.pdf},
  1998.

\bibitem{jones1976}
Eric~M Jones.
\newblock Colonization of the galaxy.
\newblock {\em Icarus}, 28(3):421--422, 1976.

\bibitem{Jones1985}
Eric~M Jones.
\newblock Where is everybody?
\newblock {\em Physics Today}, 38(8):11--13, 1985.

\bibitem{kardashev1964}
Nikolai~S Kardashev.
\newblock Transmission of information by extraterrestrial civilizations.
\newblock {\em Soviet Astronomy}, 8:217, 1964.

\bibitem{lacki2016}
Brian~C Lacki.
\newblock The log log prior for the frequency of extraterrestrial
  intelligences.
\newblock {\em arXiv preprint arXiv:1609.05931}, 2016.

\bibitem{lineweaver2001}
Charles~H Lineweaver.
\newblock An estimate of the age distribution of terrestrial planets in the
  universe: quantifying metallicity as a selection effect.
\newblock {\em Icarus}, 151(2):307--313, 2001.

\bibitem{loeb2016}
Abraham Loeb, Rafael~A Batista, and David Sloan.
\newblock Relative likelihood for life as a function of cosmic time.
\newblock {\em Journal of Cosmology and Astroparticle Physics}, 2016(08):040,
  2016.

\bibitem{merali2016}
Zeeya Merali.
\newblock Shooting for a star.
\newblock {\em Science}, 352(6289):1040--1041, 2016.

\bibitem{olson2014}
S~Jay Olson.
\newblock Homogeneous cosmology with aggressively expanding civilizations.
\newblock {\em Classical and Quantum Gravity}, 32(21):215025, 2015.

\bibitem{olson2016}
S~Jay Olson.
\newblock On the visible size and geometry of aggressively expanding
  civilizations at cosmological distances.
\newblock {\em Journal of Cosmology and Astroparticle Physics}, 2016(04):021,
  2016.

\bibitem{olson2017a}
S.~Jay Olson.
\newblock Estimates for the number of visible galaxy-spanning civilizations and
  the cosmological expansion of life.
\newblock {\em International Journal of Astrobiology}, 16(2):176--184, 004
  2017.

\bibitem{olson2017b}
S~Jay Olson.
\newblock Life-hostile conditions in the early universe can increase the
  present-day odds of observing extragalactic life.
\newblock {\em arXiv preprint arXiv:1704.04125}, 2017.

\bibitem{tegmark2014}
Max Tegmark.
\newblock {\em Our mathematical universe: My quest for the ultimate nature of
  reality}.
\newblock Vintage, 2014.

\bibitem{webb2002}
Stephen Webb.
\newblock {\em If the universe is teeming with aliens... where is everybody?:
  fifty solutions to the Fermi paradox and the problem of extraterrestrial
  life}.
\newblock Springer Science \& Business Media, 2002.

\end{thebibliography}
\bibliographystyle{plain}

\end{document}